\documentclass[pre,preprint,preprintnumbers,amsmath,amssymb,showpacs]{revtex4}
\usepackage{graphics,epsfig,epstopdf}
\newcommand{\cE}{\mathcal{E}}
\newcommand{\mV}{\mathbf{V}_0}

\newcommand{\mv}{\mathbf{\delta v}}
\newcommand{\mx}{\mathbf{x}}
\newcommand{\mE}{\mathbf{E}_0}
\newcommand{\mw}{\mathbf{V}_\bot}
\newcommand{\mk}{\mathbf{k}}
\begin {document}
\title{Linear wave action decay entailed by Landau damping in inhomogeneous, nonstationary and anisotropic plasma.}
\author{Didier B\'enisti}
\email{didier.benisti@cea.fr}
\affiliation{ CEA, DAM, DIF F-91297 Arpajon, France.}
\date{\today}
\begin{abstract}
This paper addresses the linear propagation of an electron  wave in  a collisionless, inhomogeneous, nonstationary and anisotropic plasma. The plasma is characterized by its distribution function, $f_H$, at zero order in the wave amplitude. This distribution function, from which are derived all the hydrodynamical quantities, may be chosen arbitrarily, provided that it solves Vlasov equation.  Then, from the linearized version of the electrons equation of motion, and from Gauss law, is derived an envelope equation for the wave amplitude, assumed to evolve over time and space scales much larger than the oscillation periods of the wave. The envelope equation may be cast into an equation for the  the wave action, derived from Whitham's variational principle, that demonstrates the action decay due to Landau damping. Moreover, the Landau damping rate is derived at first order in the variations of the wave number and frequency. As briefly discussed, this paper generalizes numerous previous works on the subject, provides a theoretical basis for heuristic arguments regarding the action decay, and also addresses the propagation of an externally driven wave. 
\end{abstract}
\maketitle
\section{Introduction}
The linear propagation of an electrostatic wave in an inhomogeneous and nonstationaty plasma is, clearly, an old problem in plasma physics, that has led to numerous theoretical studies for several decades (see Ref.~\cite{dodin09} and references therein for a rather exhaustive bibliography on the subject). Recently, it has attracted some renewed interest, due to its relevance to issues related to plasma compression. In particular, in Ref.~\cite{dodin09},  it is shown that  Whitham's variational principle~\cite{whitham} does apply to describe the linear propagation of an electron wave in an inhomogenous and nonstationary plasma. This result is actually not straightforward because, when the plasma density, $n$, varies in space, or in time, the Lagrangian density introduced in Whitham's theory depends on three fields, the wave amplitude, $E_p$, its eikonal, $\varphi$, and the plasma density, $n$, while the Lagrange equations are only valid for $E_p$ and $\varphi$. Therefore, it is not obvious that the variations in $n$ would entail variations in $E_p$ as predicted by Whitham's theory, i.e., such that the wave action,
\begin{equation}
\label{eq1}
\mathcal{A}Ê\equiv \int \frac{\partial \chi}{\partial \omega}E_p^2d^3\mathbf{x},
\end{equation}
be conserved. In Eq.~(\ref{eq1}), $\omega\equiv-\partial_t \varphi$ is the wave frequency, while $\chi$ is the electron susceptibility,  defined precisely by Eq.~(\ref{eq39}) of Section~\ref{III}. In the fluid limit, $\partial \chi/\partial \omega$ scales as $n^{-1/2}$, so that action conservation entails that $E_p$ should scale as $n^{3/4}$, as explained in Ref.~\cite{dodin09}. This scaling was confirmed numerically in Ref.~\cite{schmit}, where it was shown that plasma compression did induce an increase in the wave amplitude, so as to keep the wave action constant. However, it was also shown in Ref.~\cite{schmit} that compression makes the wave phase velocity decrease compared to the thermal one, which eventually entails the decay of the wave amplitude (and also of the wave action) because of Landau damping. Nevertheless, to the best of the author's knowledge, there is no unambiguous theoretical proof of the latter result. Indeed, in spite of the abundant literature on the subject, no kinetic derivation of an envelope equation, accounting for Landau damping, could be found  for an electron wave propagating in an inhomogeneous and nonstationary plasma. For example, the theoretical results of Ref.~\cite{dodin09}~relie on fluid equations and, therefore, they cannot account for Landau damping. Moreover, the expression found for $\chi$ in Eq.~(\ref{eq1}) is the fluid one, which becomes inaccurate at large temperatures. The most complete kinetic derivation of an envelope equation we are aware of, for an electron wave propagating in an inhomogeneous and non stationary plasma,  is that given in Ref.~\cite{pikulin}. This derivation stems directly from Vlasov equation, linearized about the distribution function, $f_H(\mathbf{x},\mathbf{v},t)$, at zero order in the wave amplitude. However, the results of Ref.~\cite{pikulin} lack of generality because $f_H$ obeys a force-free Vlasov equation, which is usually not the case. Moreover, and more importantly, Landau damping is not recovered in Ref.~\cite{pikulin}. An envelope equation that does account for collisionless dissipation (and, in particular, Landau damping) has been derived in Refs.~\cite{vgroup,dissip} starting from first principles. The equation found in these papers is valid in the linear and nonlinear regimes, and, in the former one,  it does show that the wave action decays at the Landau damping rate. However, the derivations of Refs.~\cite{vgroup,dissip} only hold for a homogeneous and stationary plasma. 

The present paper aims at generalizing the aforementioned articles
. The main result is Eq.~(\ref{eq74}) showing that, in the linear regime, the wave action decays at the Landau damping rate. It is derived directly from Gauss law and from the electrons equations of motion, linearized about the dynamics that makes the plasma evolve on hydrodynamical scales. This dynamics may include any kind of force field, including a magnetic one (which was not the case in previous publications). Moreover, the expression of the Landau damping rate we provide, given by Eq.~(\ref{eq46}), is valid at first order in the time and space variations of the wave number and frequency. Hence, it does not stem from a WKB approximation but vindicates, \textit{a posteriori}, this approach. 

Among all the possible applications of these results, the present work was mainly motivated by the modeling of stimulated Raman scattering (SRS). Indeed, SRS is still an issue for inertial fusion since large Raman reflectivities were measured at the Nation Ignition Facility~\cite{hinkel}. As discussed in several papers (e.g.~\cite{montgomery,strozzi}), correctly modeling collisionless dissipation is crucial to accurately predict Raman reflectivity. This might be done in an effective way by making use of envelope equations, since they proved in Refs.~\cite{srs3D,benisti10} to provide estimates for SRS reflectivity as accurate as a particle in cell or a Vlasov code, within much smaller computation times. However, the results of Refs.~\cite{srs3D,benisti10} were only for a plasma with constant density. There is, currently, no kinetic modeling of SRS in an inhomogeneous and nonstationary plasma, and the present paper is the first of a sequence of forthcoming articles that aim at filling this gap. Therefore, in order to make the application to SRS straightforward, we actually address here the propagation of a wave that may be driven. 

The derivation and presentation of our main results are organized the following way. Section~\ref{II}, provides a general expression of the charge density in terms of the plasma distribution function, $f_H$, at zero order in the wave amplitude. This charge density appears to be the sum of two contributions of distinct nature. The first contribution varies on hydrodynamical  scales, and is related to slowly-varying force field that may exist in the plasma and may make the density and temperature vary. The second contribution is at the origin of the electron plasma wave (EPW), and is used to define what may be viewed as a generalized electron susceptibility, $\Xi$. In Section~\ref{III}, $\Im(\Xi)$ is expressed in terms of the variations of the wave number, of the frequency, and of the electric field amplitude which, from Gauss law, provides the wave envelope equation in a one-dimensional geometry. This equation is first derived when there is no resonant electron in Paragraph~\ref{III.1}. In this case, it is shown to guarantee the  wave  action conservation, with $\chi$ in Eq.~(\ref{eq1}) explicitly expressed in terms of $f_H$. A wave equation that accounts for the contribution of all electrons, resonant and non resonant, is derived in Paragraph~\ref{III.2}. Then, the expression for $\chi$ is slightly changed, and, more importantly, it is shown that the action decays due to Landau damping. Moreover, our expression for the Landau damping rate accounts, at first order, for the space and time variations of the wave number and frequency. Section~\ref{IV} generalizes the results of Section~\ref{III} to a three-dimensional geometry, while Section~\ref{V} summarizes and concludes this work.


\section{The electron charge density}
\label{II}
In order to describe the propagation of an electron plasma wave that may be externally driven, one needs to calculate the charge density, $\rho_w$, induced by the EPW and the drive. As a first step of this calculation is derived, in this Section, an expression for $\rho_w$ in terms of the electron distribution function, $f_H$, at zero order in the wave and drive amplitudes. For the sake of clarity, the plasma is first assumed to be one-dimensional (1-D), and the results obtained in 1-D are then easily generalized to a three-dimensional (3-D) geometry. From $\rho_w$ is also introduced, $\Xi$, defined by Eq.~(\ref{eq13}) or  Eq.~(\ref{II6}), that may be viewed as a generalized electron susceptibility, and that allows to express Gauss law in a very simple fashion. 

\subsection{One-dimensional geometry} 
\label{II.1}
Let us denote by $-e$ the charge of an electron, by $m$, its mass, and by $f(x,v,t)$ the electron distribution function, such that
\begin{equation}
\int_{-\infty}^{+\infty} f(x,v,t)dv=n_e(x,t),
\end{equation}
where $n_e$ is the electron density. Let us, moreover, assume that there is only one ion specie (the generalization to multi-ion species is straightforward, but would require unnecessarily complicate notations), and let us denote by $n_i(x,t)$ the ion density and by $Ze$ its charge. Then, the total plasma charge density is,
\begin{equation}
\label{eq2}
\rho = -e \int_{-\infty}^{+\infty} f(x,v,t)dv +Zen_i(x,t).
\end{equation}
From Liouville theorem, the electron distribution function is conserved, so that,
\begin{equation}
\label{eq3}
f(x,v,t)=f_0[x_0(x,v,t),v_0(x,v,t)],
\end{equation}
 where $f_0$ is the initial distribution function, and $x_0$ and $v_0$ are, respectively, the initial position and velocity of an electron located at $x$, with velocity $v$, at time $t$. Let us now express $x$ and $v$ in terms of $x_0$ and $v_0$ the following way,
\begin{eqnarray}
\nonumber
x&=&x_0+v_0t+\delta x_H+\delta x \\
\label{eq4}
&\equiv & X_0+\delta x \\
\nonumber
v&=&v_0+\delta v_H+\delta v \\
\label{eq5}
&\equiv& V_0+\delta v,
\end{eqnarray}
where $\delta x$ and $\delta v$ are induced by the EPW and the drive, while $\delta x_H$ and $\delta v_H$ evolve on hydrodynamical scales. The two latter quantities are specifically defined by, $d\delta_H/dt=\delta v_H$ and $d\delta v_H/dt=F_H/m$, where $F_H$ is the force that makes all the hydrodynamical quantities, such as he local density and temperature, vary.  More precisely, the distribution function, $f_H(X_0,V_0,t)$, at zero order in the wave and drive amplitudes, satisfies the following Vlasov equation,
\begin{equation}
\label{eq6}
\frac{\partial f_H}{\partial t}+V_0\frac{\partial f_H}{\partial X_0}+\frac{F_H}{m}\frac{\partial f_H}{\partial V_0}=0.
\end{equation}
Then, if at time $t=0$, the wave and drive amplitudes are vanishingly small, Eq.~(\ref{eq6}) translates into,
\begin{equation}
\label{eq7}
f_H(X_0,V_0,t)=f_0(X_0-v_0t-\delta x_H,V_0-\delta v_H).
\end{equation}
From Eq.~(\ref{eq3}), and the definitions~Eqs.~(\ref{eq4})~and~(\ref{eq5})~of $X_0$ and $V_0$, Eq.~(\ref{eq7}) is equivalent to,
\begin{equation}
\label{eq8}
f(x,v,t)=f_H(X_0,V_0,t). 
\end{equation}
Plugging Eq.~(\ref{eq8})~into~Eq.~(\ref{eq2}) for the charge density, and making use of the change of variables $v\rightarrow V_0$ in the integral of Eq.~(\ref{eq2}), yields,
\begin{equation}
\label{eq9}
\rho= -e \int_{-\infty}^{+\infty} f_H(X_0,V_0,t)dV_0 +Zen_i(x,t)- e\int_{-\infty}^{+\infty} f_H(X_0,V_0,t) \frac{\partial\delta v}{\partial V_0}dV_0.
\end{equation}
Since we aim at deriving an envelope equation for the EPW that is only  at first order in the variations of the density, we now use $f_H(X_0,V_0,t)\approx f_H(x,V_0,t)-\delta x \partial_xf_H(x,V_0,t)$. Note that latter approximation is \textit{not} a linearization with respect to the wave amplitude, but an expansion at first order in the space variations of $f_H$. Then,
\begin{eqnarray}
\nonumber
\rho&\approx& -e \int_{-\infty}^{+\infty} f_H(x,V_0,t)dV_0 +Zen_i(x,t)+e \int_{-\infty}^{+\infty} \delta x \partial_xf_H(x,V_0,t)dV_0\\
\label{eq10}
&&- e\int_{-\infty}^{+\infty} f_H(x,V_0,t) \frac{\partial\delta v}{\partial V_0}dV_0-e\int_{-\infty}^{+\infty} \partial_xf_H(x,V_0,t) \delta x\frac{\partial\delta v}{\partial V_0}dV_0\\
&\equiv& \rho_H+\rho_w,
\end{eqnarray}
where
\begin{equation}
\label{eq11}
\rho_H = -e \int_{-\infty}^{+\infty} f_H(x,V_0,t)dV_0 +Zen_i(x,t),
\end{equation}
evolves on hydrodynamical scales, and is considered as a given quantity. As for the charge density, $\rho_w$, induced by the wave and the external drive, from now on, it is identified with its linear value. Hence, by making use of an integration by part for the third term of Eq.~(\ref{eq10}), one finds,
\begin{equation}
\label{eq12}
\rho_w(x,t) \approx e \int_{-\infty}^{+\infty} \delta x \partial_x f_H(x,V_0,t)dV_0+ e\int_{-\infty}^{+\infty}\delta v  f'_H(x,V_0,t)dV_0,
\end{equation}
where we have denoted, $f'_H\equiv \partial_{V_0}f_H$. 

We now assume that the electric field of the EPW and the drive writes,
\begin{equation}
\label{eq12a}
E=E_pe^{i\phi}+iE_de^{i(\phi+\delta \phi)}+c.c.,
\end{equation}
where $c.c.$ stands for the complex conjugate, and where $E_p$ and $E_d$  are, respectively, the \textit{real} amplitudes of the EPW and drive fields, that vary much more slowly than the eikonal, $\phi$. From $\phi$ are defined the wave number, $k\equiv \partial_x\phi$ and wave frequency, $\omega\equiv-\partial_t\phi$. Both $E_p$ and $E_d$ are at the origin of the charge density, $\rho_w$, so that $E$ will henceforth be termed the~\textit{total}~electric field, although the force $F_H$ may include a slowly varying electric field. Clearly, $E$, may also be written the following way,
\begin{equation}
\label{eq12b}
E=E_0e^{i\varphi}+c.c.,
\end{equation}
where $E_0 \equiv \sqrt{E_p^2+E_d^2-2E_dE_p\sin(\delta \phi)}$ is the total field amplitude (which is clearly real), and $e^{i\varphi} \equiv E_0^{-1}e^{i\phi}(E_p+iE_de^{i\delta \phi})$. We moreover assume that the wave is nearly on resonance with the drive, so that $k\approx \partial_x\varphi$ and $\omega\approx-\partial_t\varphi$. 

Similarly, we write the charge density, $\rho_w$, as,
\begin{equation}
\label{eq13a}
\rho_w=\rho_0e^{i\varphi}+c.c.,
\end{equation}
and introduce,
\begin{equation}
\label{eq13}
\Xi \equiv \frac{i\rho_0}{\varepsilon_0kE_0},
\end{equation}
which may be viewed as a generalized electron susceptibility. 

In the total field $E$, only the EPW electric field is self-consistently related to $\rho_w$, since the drive is imposed externally (for example, in the case of SRS, the drive is due to electromagnetic laser fields, and is just the component of the ponderomotive force along the wave number). Then, Gauss law is just $\partial_x(E_pe^{i\phi}+c.c.)=\rho_w/\varepsilon_0$, which reads,
\begin{eqnarray}
\nonumber
(ikE_p+\partial_xE_p)e^{i\phi}&=&-ik\Xi E_0e^{i\varphi}\\
\label{eq14}
&=&-ik\Xi(E_p+iE_de^{i\delta \phi})e^{i\phi}.
\end{eqnarray}
Since $E_0$, $E_p$ and $E_d$ are real, Eq.~(\ref{eq14}) is equivalent to the two following equations, 
\begin{eqnarray}
\label{eq15a}
(1+\Xi_r)E_p&=&E_d[\Xi_i\cos(\delta \phi)+\Xi_r\sin(\delta \phi)], \\
\label{eq16a}
k^{-1}\partial_xE_p&=&\Xi_iE_p+E_d[\Xi_r\cos(\delta \phi)-\Xi_i\sin(\delta \phi)],
\end{eqnarray}
where $\Xi_r\equiv\Re(\Xi)$ and $\Xi_i=\Im(\Xi)$. We henceforth assume that $E_d \ll E_p$ which, as shown in Ref.~\cite{benisti08}, is the case for SRS. Then, Eq.~(\ref{eq15a}) is approximated by
\begin{equation}
\label{eq15}
1+\Xi_r=0,
\end{equation}
which actually yields the EPW dispersion relation. The validity of the approximation that leads to Eq.~(\ref{eq15}) is discussed in detail in Ref.~\cite{benisti08}, where it is shown that accounting for the drive amplitude in the dispersion relation is important to calculate the nonlinear frequency shift of the plasma wave, but not to correctly estimate the linear frequency. Since we restrict here to linear wave propagation, Eq.~(\ref{eq15}) may be considered as exact. 

As will be shown in Sections~\ref{III} and \ref{IV}, $\Xi_i$ is of the order of $(kL_0)^{-1}$ or $(\omega T_0)^{-1}$, where $L_0$ and $T_0$ are, respectively, the typical length and time scales of variation of $E_0$, $k$ or $\omega$. Hence, for the slow variations considered in this paper, $\Xi_i \ll 1$, so that, from Eq.~(\ref{eq15}), $\Xi_i\ll \Xi_r$. This lets us conclude that Eq.~(\ref{eq16a}) may be approximated by,
\begin{equation}
\label{eq16}
\Xi_iE_p-k^{-1}\partial_xE_p=E_d \cos(\delta \phi).
\end{equation}
  
\subsection{Three-dimensional geometry} 
\label{II.2}
The results obtained in Paragraph~\ref{II.1} straightforwardly generalize to a three-dimensional geometry. Just like for a 1-D plasma, let us introduce the distribution function, $f_H(\mathbf{X}_0, \mathbf{V}_0,t)$, at zero order in the wave amplitude, that obeys the following Vlasov equation,
\begin{equation}
\label{II1}
\frac{\partial f_H}{\partial t}+\mathbf{V}_0.\frac{\partial f_H}{\partial \mathbf{X}_0}+\frac{\mathbf{F}_H}{m}.\frac{\partial f_H}{\partial \mathbf{V}_0}=0.
\end{equation}
Then, Eq.~(\ref{eq9}) for the charge density becomes,
\begin{equation}
\label{II2}
\rho= -e \iiint f_H(\mathbf{X}_0,\mathbf{V}_0,t) \left \vert \frac{\partial (\mathbf{V}_0+\mathbf{\delta v})}{\partial \mathbf{V}_0} \right \vert d\mV+Zen_i(x,t),
\end{equation}
where $\vert \partial (\mV+\mv)/\partial \mV \vert$ is the Jacobian of the change of variables $\mV+\mv \rightarrow \mV$. The linear value of this Jacobian is, 
\begin{equation}
\label{II3}
 \left \vert \frac{\partial (\mathbf{V}_0+\mathbf{\delta v})}{\partial \mathbf{V}_0} \right \vert_{lin} = 1+ \sum_n \frac{\partial \delta v_n}{\partial V_{0n}}, 
\end{equation}
where $\delta v_n$ and $V_{0n}$ are, respectively, the $n^{th}$ component of $\mv$ and $\mV$. Therefore, following the lines of Paragraph~\ref{II.1}, one straightforwardly finds that the linerarized charge density induced by the wave and the drive is,
\begin{equation}
\label{II4}
\rho_w(\mx,t) = e \iiint\delta \mx.\partial_\mx f_H(\mx,\mV,t)d\mV+ e\iiint \mv .\partial_{\mV}f_H(\mx,\mV,t)d\mV.
\end{equation}
Let us now write $\rho_w$ and the total electric field as in Paragraph~\ref{II.1},
\begin{eqnarray}
\label{II5}
\rho_w &=& \rho_0e^{i\varphi}+c.c.,\\
\mathbf{E}&=& \mathbf{E}_pe^{i\phi}+\mathbf{E}_de^{i(\phi+\delta \phi)}+c.c.\\   
&\equiv &\mE e^{i\varphi}+c.c.,
\end{eqnarray}
where  $\mathbf{E}_p$, $\mathbf{E}_d$ and $\mE$ are real vectors, and let us introduce,
\begin{equation}
\label{II6}
\Xi \equiv \frac{i\rho_0}{\varepsilon_0 \mathbf{k}.\mE}.
\end{equation}
Then, with the same approximations as in Paragraph~\ref{II.1}, Gauss law yields the following equations, 
\begin{eqnarray}
\label{II7}
1+\Xi_r&=&0, \\
\label{II8}
\Xi_i-\frac{\mathbf{\nabla}.\mathbf{E}_p}{\mathbf{k}.\mathbf{E}_p}&=&\frac{\mk.\mathbf{E}_d }{\mk.\mathbf{E}_p}\cos(\delta \phi). 
\end{eqnarray}

\section{Wave equation in a one-dimensional geometry}
\label{III}
\subsection{Plasma with no resonant electron}
\label{III.1}
For the sake of clarity, we first derive $\Xi_i$ in a one-dimensional geometry. Moreover, we start by assuming that there is no resonant electron, the exact meaning of this assumption being clarified after Eq.~(\ref{eq25}) and at the end of Subsection \ref{III.2.2}. 

From Eq.~(\ref{eq12}), we now need to calculate $\delta x$ and $\delta v$ in order to derive $\Xi_i$. This is done by solving the electrons equations of motion,
\begin{eqnarray}
\label{eq17}
\frac{d\delta x}{dx}&=&\delta v, \\
\label{eq18}
\frac{d\delta v}{dt}&=&-\frac{e}{m}E_0^{i\varphi}+c.c.
\end{eqnarray}
Eq.~(\ref{eq18}) is solved by making use of an integration by parts, which yields, at first order in the variations of $E_0$, 
\begin{equation}
\label{eq19}
\frac{-m}{e}\delta v \approx  \cE_0(t)I_1(t)-\left.\frac{d\cE_0}{dt'}\right \vert_{t'=t}I_2(t)+c.c.,
\end{equation}
where $\cE_0(t) \equiv E_0[x(t),t]$, and 
\begin{equation}
I_1(t) \equiv \int \cE_0(t') e^{i\varphi[x(t'),t']} dt'
\end{equation}
is one primitive of $\cE_0e^{i\varphi}$, and
\begin{equation}
I_2(t') \equiv \int I_1(t')dt'
\end{equation}
is one primitive of $I_1$. Since we only look for a first order envelope equation, we now express $I_1$ at first order in the variations of $k$ and $\omega$, and $I_2$ at zero order in these variations. Then, one easily finds,
\begin{eqnarray}
\label{eq20}
I_1(t')& \approx& e^{i\psi(t')}\left[ \frac{-i }{d\psi/dt'}-\frac{d^2\psi/dt'^2}{(d\psi/dt')^3} \right],\\
\label{eq20b}
 I_2(t')& \approx&- \frac{e^{i\psi(t')}}{(d\psi/dt')^2},
\end{eqnarray}
where we have denoted $\psi(t')\equiv \varphi[x(t'),t']$. 

Let us now provide an expression for $\delta v$ that is linearized with respect to $E_0$, the zero-order motion being defined by the hydrodynamical force, $F_H$. To do so, we make use of the approximations, $\cE_0(t') \approx E_0[x-\delta X_0(t'),t']$ and $\psi(t') \approx \varphi [x-\delta X_0(t'),t']$, where $\delta X_0(t') \equiv \int_{t'}^t V_0(t'')dt''$ with $dV_0/dt=F_H/m$. This yields, 
\begin{eqnarray}
\label{eq21}
d\cE_0/dt'\vert_{t'=t}&=&\partial_tE_0+V_0\partial_xE_0,\\
\label{eq22}
d\psi/dt'\vert_{t'=t}&=&-\omega+kV_0,\\
\label{eq22}
d^2\psi/dt'^2\vert_{t'=t}&=& -\partial_{t} \omega+2V_0 \partial_t k+V_0^2\partial ^2_x k+kF_H/m, 
\end{eqnarray}
where, in Eq.~(\ref{eq22}), we used $\partial_tk=-\partial_x\omega$ and $dV_0/dt=F_H/m$. From Eqs.~(\ref{eq20}-\ref{eq22}),  one easily finds,
\begin{eqnarray}
\label{eq23}
\nonumber
I_1(t)& \approx& e^{i\varphi(x,t)}\left[ \frac{-i }{kV_0-\omega}+\frac{\partial_t \omega-kF_H/m-2V_0 \partial_t k/\partial t-V_0^2 \partial^2k/\partial x^2}{(kV_0-\omega)^3} \right]\\
\\
\label{eq24}
 I_2(t)& \approx&- \frac{e^{i\varphi(x,t)}}{(kV_0-\omega)^2}. 
\end{eqnarray}
Plugging Eqs.~(\ref{eq23}) andÊ(\ref{eq24}) into the expression~(\ref{eq19})~for $\delta v$ yields, $\delta v=\delta v_0 e^{i\varphi}+c.c.$, with
\begin{equation}
\label{eq25}
-\frac{m}{e}\delta v_0 \approx  \frac{-i E_0}{kV_0-\omega}+\frac{\partial_tE_0+V_0\partial_xE_0}{(kV_0-\omega)^2}+E_0\frac{\partial_t \omega-kF_H/m-2V_0 \partial_t k-V_0^2 \partial_xk}{(kV_0-\omega)^3}.
\end{equation}
Note that our expansions, leading to Eq.~(\ref{eq25}) for $\delta v$, only make sense if, whatever $V_0$, $(\omega-kV_0) T_0\gg 1$ and $(\omega/V_0-k) L_0 \gg 1$, where $T_0$ and $L_0$ are, respectively, the typical time and length scales of variation of $E_0$, $k$ or $\omega$. The latter conditions are precisely what we mean by the conditions for no resonant electron in the plasma. 

Since, in the expression Eq.~(\ref{eq12}) for $\rho_w$, $\delta x$ is multiplied by $\partial_x f_H$, it only needs to be estimated at zero-order in the variations of $E_0$, $k$ and $\omega$. Then, it is easily found that $\delta x =\delta x_0e^{i\varphi}+c.c.$ with,
\begin{equation}
\label{eq26}
-\frac{m}{e}\delta x_0Ê\approx \frac{-E_0}{(kV_0-\omega)^2}.
\end{equation}
From the knowledge of $\delta v$ and $\delta x$ given by Eqs.~(\ref{eq25}) and (\ref{eq26}), $\Xi$ is easily derived by making use of the expression Eq.~(\ref{eq12}) for $\rho_w$ and of the relation Eqs.~(\ref{eq13a}) and (\ref{eq13}) between $\rho_w$ and $\Xi$. The corresponding calculations are detailed in the Appendix~\ref{A} where $\Xi$ is expressed in terms of the electron susceptibility, $\chi$, defined here as,
\begin{equation}
\label{eq27}
\chi = -\frac{e^2}{\varepsilon_0mk} \int_{-\infty}^{+\infty}Ê\frac{f'_H}{kV_0-\omega}dV_0.
\end{equation}
It is then found that $\Xi_r=\chi$, so that Eq.~(\ref{eq14}) reads
\begin{equation}
\label{eq28}
1+\chi=0.
\end{equation}
As for $\Xi_i$, we find,
\begin{equation}
\label{eq29}
\Xi_i=\frac{1}{2E_0^2}\left[\frac{\partial}{\partial t}\left( \frac{\partial \chi}{\partial \omega}E_0^2\right)-\frac{\partial}{\partial x}\left( \frac{\partial \chi}{\partial k}E_0^2\right) \right]-\frac{\chi}{kE_0}\frac{\partial E_0}{\partial x}-\frac{1}{k}\frac{\partial \chi}{\partial x}. 
\end{equation}
From Eq.~(\ref{eq28}), $1+\chi=0$ whatever $x$, so that $\partial_x\chi=0$. Moreover, we henceforth use the condition $E_d \ll E_p$ to replace, in the expression of $\Xi_i$, $E_0$ by $E_p$. This actually amounts to deriving an envelope equation at zero order in the variations of the drive amplitude, which is known to be enough to correctly address SRS~\cite{srs3D}.  Then, Eq.~(\ref{eq16}) simply reads
\begin{equation}
\label{eq30}
\frac{\partial}{\partial t}\left( \frac{\partial \chi}{\partial \omega}E_p^2\right)-\frac{\partial}{\partial x}\left( \frac{\partial \chi}{\partial k}E_p^2\right) =2E_pE_d\cos(\delta \varphi).
\end{equation}
When $E_d=0$, Eq.~(\ref{eq30}) is exactly the equation one would derive from Whitham's variational principle, that guarantees the conservation of the wave action defined by Eq.~(\ref{eq1}).

\subsection{Plasma with all electrons}
\label{III.2}
Let us now address the general situation when the plasma is composed on two kinds of electrons, the resonant and the non-resonant ones. In Paragraph~\ref{II.1}, the non-resonant electrons were defined by the condition, $\vert V_0-v_\phi \vert >\Delta V_R$, with $k\Delta V_RT_0\gg 1$ and $(\Delta V_R/v_\phi)kL_0\gg 1$. In the remainder of this paper, we use a more specific definition. Indeed, all the calculations are henceforth performed by assuming that, for each local value of the phase velocity, $v_\phi$, $V_0$ for the resonant electrons exactly varies between $v_\phi-\Delta V_R$ and $v_\phi+\Delta V_R$ (while $V_0$ for the non-resonant electrons lies in the complementary interval). However, there is a technical difficulty related to this local definition for the resonant and non-resonant electrons, that is discussed in the Appendix~\ref{C}. As shown in this Appendix, this difficulty may actually be ignored (as will be done here) without affecting the wave equation.
 
\subsubsection{Basic assumption on the distribution function}
\label{III.2.1}

We, henceforth, assume that the resonance width $\Delta V_R$ is such that, $\Delta V_R \ll v_T$, where $v_T$ is the typical scale of variation, in velocity, of $f'_H$ about $v_\phi$. For a Maxwellian, $v_T$ would just be the electron thermal velocity. For any bell-shaped distribution function, centered at $V_0 \approx 0$, it is clear that there is a significant fraction of resonant electrons only if $v_T$ is sufficiently large compared to $v_\phi - \Delta V_R$. Hence, it is quite natural to assume $\Delta V_R \ll v_T$ when the contribution to the charge density from the resonant electrons does matter. One noticeable exception to the previous rule is the situation when there is a very cold electron beam, with mean velocity close to $v_\phi$, in the plasma. Clearly, the present paper cannot address that situation. However, a very detailed study of the cold beam-plasma instability, in the linear and non linear regimes, will be the subject of forthcoming articles.

Therefore, we shall henceforth assume $\Delta V_R \ll v_T$, and express $\Xi$ as the sum of two contributions, 
\begin{equation}
\label{eq31}
\Xi \equiv \Xi^{res}+\Xi^{nres},
\end{equation}
where $\Xi^{res}$ and $\Xi^{nres}$ are, respectively, from resonant and non-resonant electrons.

\subsubsection{Contribution from the resonant electrons and Landau damping}
\label{III.2.2}
The charge density, $\rho_w^{res}$, induced by the resonant electrons is given by Eq.~(\ref{eq12}), with $f_H$ assumed to be arbitrarily small when $\vert V_0-v_\phi \vert > \Delta V_R$. Because of the hypothesis, $\Delta V_R \ll v_T$, to leading order, 
\begin{eqnarray}
\nonumber
\rho_w^{res}(x,t) \approx &&e \int_{v_\phi-\Delta V_R}^{v_\phi-\Delta V_R} \delta x \partial_x \left[f_H(x,v_\phi,t)+(V_0-v_\phi)f'_H(x,v_\phi,t)\right]dV_0\\
\label{eq32}
&&+ e\int_{v_\phi-\Delta V_R}^{v_\phi+\Delta V_R}\delta v  f'_H(x,v_\phi,t)dV_0.
\end{eqnarray}

For the sake of clarity, the detailed derivation of $\rho_w^{res}$, accounting for the slow variation of $k$ and $\omega$, is reported in the Appendix \ref{B}. In this Subsection, we shall restrict to a homogenous and stationary plasma, in order to discuss in a very simple fashion the derivation of Landau damping. For a homogeneous and stationary plasma, 
\begin{equation}
\rho_w^{res}(x,t) \approx  e\int_{v_\phi-\Delta V_R}^{v_\phi+\Delta V_R}\delta v  f'_H(x,v_\phi,t)dV_0,
\end{equation}
so that $\rho_w^{res}(x,t) \equiv \rho_0e^{i(kx-\omega t)}+c.c.$, with,
\begin{equation}
\label{eq32.1}
\rho_0=-\frac{e^2}{m}f'_H(x,v_\phi,t)\int_0^t\int_{v_\phi-\Delta V_R}^{v_\phi+\Delta V_R} E_0[x(t'),t']e^{i(kV_0-\omega)(t'-t)}dV_0dt'.
\end{equation}
We now express $E_0(x,t)$ in terms of a Fourier integral, $E_0(x,t)Ê\equiv \int_{-\infty}^{+\infty} \tilde{E}_\kappa(t)e^{i\kappa x}d\kappa$, to find
\begin{equation}
\label{eq32.2}
\rho_0 \approx -\frac{e^2}{m}f'_H(x,v_\phi,t)  \int_0^t\int_{-\infty}^{+\infty}  \tilde{E}_\kappa(t')e^{i\kappa x}\int_{v_\phi-\Delta V_R}^{v_\phi+\Delta V_R} e^{i[(k+\kappa)V_0-\omega](t'-t)}dV_0d\kappa dt'.
\end{equation}
We, now, make use of the change of variables, $u\equiv (k+\kappa)\Delta V_R(t'-t)$, in the time integral of Eq.~(\ref{eq32.2}), to find,
\begin{equation}
\label{eq33}
\rho_0 \approx  -\frac{e^2}{m}f'_H(x,v_\phi,t)\int_{-k\Delta V_Rt}^0 \int_{-\infty}^{+\infty} \frac{2e^{i\kappa x}}{k+\kappa}\tilde{E}_\kappa[t+u/(k+\kappa)\Delta V_R]\frac{\sin u}{u} d\kappa du.
\end{equation}
Now, $E_0$ is supposed to vary very little over one wavelength, so that $\tilde{E}_\kappa$ is only significant when $\kappa \ll k$, and one may assume $(k+\kappa)\Delta V_R \approx k \Delta V_R$. Moreover, it is well know that $\sin u/u$ is a function whose width is of the order of $2\pi$. Hence, the integrand in Eq.~(\ref{eq33}) assumes significant values only when $u/(k+\kappa)\Delta V_R \alt 2\pi/k\Delta V_R$. Since, by definition, $1/k\Delta V_R$ is much less than the typical time of variation, $T_0$, of $E_0$, we conclude that $\tilde{E}_\kappa[t+u/(k+\kappa)\Delta V_R]$ may be replaced by $\tilde{E}_\kappa(t)$. Hence, at first order in the space variations of $E_0$, 
\begin{eqnarray}
\nonumber
\int_{-\infty}^{+\infty} \frac{e^{i\kappa x}}{k+\kappa}\tilde{E}_\kappa[t+u/(k+\kappa)\Delta V_R]d\kappa \approx &&\int_{-\infty}^{+\infty} \frac{\tilde{E}_\kappa(t)}{k+\kappa} e^{i\kappa x}d\kappa \\
\nonumber
\approx && \int_{-\infty}^{+\infty} \left(1-\frac{\kappa}{k}Ê\right)\tilde{E}_\kappa(t) e^{i\kappa x}d\kappa \\
\label{eq32.3}
= && \frac{E_0(x,t)}{k}+\frac{i}{k^2}\frac{\partial E_0}{\partial x}.
\end{eqnarray}
Hence, from Eq.~(\ref{eq33}), we find, 
\begin{equation}
\label{eq32.4}
\rho_0 \approx -\frac{2e^2}{mk}f'_H(x,v_\phi,t)e^{i(kx-\omega t)}\left[E_0(x,t)+\frac{i}{k}\frac{\partial E_0}{\partial x}\right]\int_{-k\Delta V_Rt}^0 \frac{\sin u}{u} du.
\end{equation}
We shall now restrict to times, $t$, such that $k\Delta V_R t \gg 1$, a condition that will be discussed in great detail in a few lines. Then, in the integral of Eq.~(\ref{eq33}),  $-k\Delta V_R t$ may be replaced by $-\infty$, so that
\begin{equation}
\label{eq34}
\rho_0 \approx -\frac{\pi e^2}{mk}f'_H(x,v_\phi,t)\left[E_0(x,t)+\frac{i}{k}\frac{\partial E_0}{\partial x}\right].
\end{equation}
From Eq.~(\ref{eq34}) and the definition Eq.~(\ref{eq13a}) of $\Xi$, we conclude that, for a homogenous and stationary plasma,
\begin{equation}
\label{eq35}
\Xi^{res} \approx -\frac{iÊ\pi e^2}{\varepsilon_0m k^2}f'_H(x,v_\phi,t)\left[1+\frac{i}{kE_0}\frac{\partial E_0}{\partial x}\right].
\end{equation}
As is quite obvious, and will be shown in Subsection~\ref{III.2.4}, this term, combined with the contribution from the non-resonant electrons, will provide the Landau damping of the wave action. 

Let us now discuss physically the condition $k\Delta V_R t \gg 1$ (in practice $k\Delta V_R t > 2\pi$ is enough), we had to impose in order to approximate $\rho_0$ by the expression given in Eq.~(\ref{eq34}). As we shall now show it, this condition is actually not restrictive. Indeed, we only discuss the propagation of an EPW once its amplitude is much larger than the noise level (otherwise, it makes non sense to assume slow variations for $E_0$). Therefore, this EPW has necessarily grown, either by itself or due to some external drive, at least during a finite time.  Hence, if we denote by $t_g$ the typical time of variation of the EPW during its growth, we only consider times $t \gg t_g$. For a wave that keeps growing, $t_g$ is the same as $T_0$, the typical timescale of variation of $E_0$. In this case, the condition $k\Delta V_R t \gg 1$ is necessarily fulfilled since, by definition, $k\Delta V_R T_0 \gg 1$. However, one may also consider the situation when, after having been driven for a finite time, the EPW experiences Landau damping. Then, during the damping phase, $T_0 \sim 1/\nu_L$, where $\nu_L$ is the Landau damping rate. In this case, $\Delta V_R$ is such that $k\Delta V_R/\nu_L \gg 1$, and Eq.~(\ref{eq34}) for $\rho_0$ is only accurate after a time $t$ such that $k\Delta V_R t \agt 2\pi$. Hence, our expression for $\Xi^{res}$, and therefore our derivation of Landau damping, is only valid after a time that scales as $1/\nu_L$, although it may be small compared to $1/\nu_L$. Now, it is well known that there exists a finite time before the settlement of Landau damping, because Landau damping is only an asymptotic result. We, therefore, conclude that the condition  $k\Delta V_R t \gg 1$, which we had to impose in order to obtain the expression Eq.~(\ref{eq35}) for $\Xi^{res}$, is not restrictive. 

Moreoever, it is quite interesting to discuss further, in terms of $\Delta V_R$, the conditions under which the value of $\rho_0$, given by Eq.~(\ref{eq33}), is close to its approximate expression, Eq.~(\ref{eq34}). There are basically two conditions. The first one is that the integral, $\int_{-k\Delta V_R t}^0 \sin u/u du$, has to be close to $\pi/2$, which requires a resonance width, $\Delta V_R \agt 2\pi/kt$. The second one is that $\tilde{E}_\kappa(t+u/k\Delta V_R) \approx \tilde{E}_\kappa(t)$, which requires a resonance width that has to be, at least, of the order of $1/T_0$. When $T_0 \sim 1/\nu_L$, we recover the result of Refs.~\cite{drummond,gris} that the resonance width for Landau damping first decreases as $1/t$, until it saturates to a value of the order of $\nu_L/k$. Here, this result is generalized to a wave that may grow or decay, and whose amplitude may vary in space and time.

The initial decrease of the resonance width, $\Delta V_R\sim 2\pi/kt$, has a very simple physical meaning, that we shall now explain. We define $\Delta V_R$ such that, when $\vert V_0 -v_\phi \vert >\Delta V_R$, $\delta v$ may be accurately approximated by Eq.~(\ref{eq25}). This approximation does not only require a nearly linear electron motion but, also, a nearly adiabatic one. Indeed, the first term in Eq.~(\ref{eq25}) is at zero order in the space and time variations of the fields and, actually, it is easily found that this term is only the linear limit of  the adiabatic value of $\Xi$ derived in Ref.~\cite{benisti07}. The next two terms of Eq.~(\ref{eq25}) may be simply viewed as first order terms in an expansion about a purely adiabatic motion. Hence, the non-resonant electrons are the nearly adiabatic ones. Now, as is well known, only after a finite time, $t_a$, may one neglect the non-adiabatic part of the electron response. Here, we just find that  $t_a \sim 2\pi/k\Delta V_R$, that is the the period of an electron orbit in the wave frame (provided that $V_0$ is very different from $v_\phi$).

Since the non-resonant electrons are the nearly adiabatic ones, we conclude from Section~\ref{III.1} that, when all electrons are nearly adiabatic, the wave action is conserved, as predicted by Whitham theory. Hence, the conservation of the wave action, defined by Eq.~(\ref{eq1}), is closely related to the conservation of the electron dynamical action, $I\equiv (k/2\pi)\int_0^{2\pi/k}vdx$. This is not surprising since Whitham theory introduces a local Lagrangian density, which, therefore, assumes a local electron response, that is only true if the electrons are adiabatic. \\

In order to conclude this Subsection, let us now report the results of the Appendix~\ref{B}, where $\Xi^{res}$ is derived for a nonstationary and inhomogeneous plasma. In particular, account is taken of the space and time variations of the wave number and frequency. Then, Eq.~(\ref{eq35}) is changed into,
\begin{equation}
\label{eq36}
\Xi^{res} \approx -i\left[1-\frac{i}{k^2}Ê\frac{\partial k}{\partial x}+\frac{i}{k}Ê\frac{\partial E_0}{\partial x}\right] \frac{\pi e^2}{\varepsilon_0mk^2}f'_H(x,v_\phi,t).
\end{equation}

\subsubsection{Contribution from the non-resonant electrons}
\label{III.2.3}
The contribution, $\Xi^{nres}$, from the non-resonant electrons, has already been derived in Section~\ref{III.1}. Using the results of this Section, we introduce,
\begin{equation}
\label{eq38}
\chi = -\frac{e^2}{\varepsilon_0mk} \int_{-\infty}^{+\infty}Ê\frac{f^{'nres}_H}{kV_0-\omega}dV_0,
\end{equation}
where $f^{nres}_H$ is the distribution of the non-resonant electrons, $f_H^{nres}(x,V_0,t)Ê\approx 0$ when $\vert V_0-v_\phi \vert \leq \Delta V_R$. Since $\Delta V_R$ is assumed to be much less than the typical scale, $v_T$, of variation of $f'_H$, one may approximate $\chi$ by,
\begin{equation}
\label{eq39}
\chi \approx -\frac{e^2}{\varepsilon_0mk} P.P. \left(\intÊ\frac{f'_H}{kV_0-\omega}dV_0\right),
\end{equation}
where $f_H$ now accounts for all electrons, and where $P.P.$ stands for the Cauchy principal part. Now that $\chi$ is defined, one may use the results of Section~\ref{III.1} for  the real part of $\Xi^{nres}$,
\begin{equation}
\label{eq40}
\Xi^{nres}_r=\chi, 
\end{equation}
and the expression Eq.~(\ref{eq29}) for its imaginary part, 
\begin{equation}
\label{eq41}
\Xi_i^{nres}=\frac{1}{2E_0^2}\left[\frac{\partial}{\partial t}\left( \frac{\partial \chi}{\partial \omega}E_0^2\right)-\frac{\partial}{\partial x}\left( \frac{\partial \chi}{\partial k}E_0^2\right) \right]-\frac{\chi}{kE_0}\frac{\partial E_0}{\partial x}-\frac{1}{k}\frac{\partial \chi}{\partial x}.
\end{equation}
Using Eqs.~(\ref{eq36})~and~(\ref{eq40}), one finds that the equation $1+\Xi_r=0$ [i.e., Eq.~(\ref{eq15}) of Paragraph~\ref{II.1} deduced from Gauss law] reads,
\begin{equation}
\label{eq43}
\chi=-1+\frac{\pi e^2}{\varepsilon_0m k^2}f'_H(x,v_\phi,t)\left[-\frac{1}{k^2}\frac{\partial k}{\partial x}+\frac{1}{kE_0}\frac{\partial E_0}{\partial x}\right].
\end{equation}
Hence, up to terms of order two in the space variations of the fields, 
\begin{equation}
\label{eq44}
\Xi_i^{nres}\approx \frac{1}{2E_0^2}\left[\frac{\partial}{\partial t}\left( \frac{\partial \chi}{\partial \omega}E_0^2\right)-\frac{\partial}{\partial x}\left( \frac{\partial \chi}{\partial k}E_0^2\right) \right]+\frac{1}{kE_0}\frac{\partial E_0}{\partial x}.
\end{equation}

\subsubsection{Landau damping of the wave action}
\label{III.2.4}
In order to find the equation ruling the evolution of the wave action, we first need to use the same approximation as in Paragraph~\ref{II.1} that, in the expression for $\Xi_i$ deduced from Eqs.~(\ref{eq36})~and~(\ref{eq44}), one may replace $E_0$ by $E_p$. Then Eq.~(\ref{eq16}), $\Xi_iE_p-k^{-1}\partial_xE_p=E_d \cos(\delta \phi)$, yields, 
\begin{equation}
\label{eq45}
\frac{\partial}{\partial t}\left( \frac{\partial \chi}{\partial \omega}E_p^2\right)-\frac{\partial}{\partial x}\left( \frac{\partial \chi}{\partial k}E_p^2\right) +2\nu_L \frac{\partial \chi}{\partial \omega}E_p^2=2E_pE_d\cos(\delta \phi),
\end{equation}
where
\begin{equation}
\label{eq46}
\nu_L \equiv -\frac{\pi e^2}{\varepsilon_0m k^2Ê\partial_\omega \chi}f'_H(x,v_\phi,t),
\end{equation}
is the Landau damping rate. It is noteworthy that the expression for $\nu_L$, at first order in the variations of $k$ and $\omega$, is exactly the same as for a homogeneous and stationary plasma, which vindicates a WKB approach for Landau damping.

With our definition, $f_H$ is proportional to the local electron density, $n_e(x,t)$, and one may want to introduce the local plasma frequency, $\omega_{pe}^2=n_ee^2/\varepsilon_0m$, in order to provide a more familiar expression for $\nu_L$, 
\begin{equation}
\label{eq47}
\nu_L \equiv -\frac{\pi \omega_{pe}^2}{ k^2Ê\partial_\omega \chi}\frac{f'_H(x,v_\phi,t)}{n_e(x,t)}.
\end{equation}

Clearly, Eq.~(\ref{eq45}) shows that, if the wave is not driven, its action decreases at twice the Landau damping rate. 
\section{Wave equation in a three-dimensional geometry}
\label{IV}
The derivation of the wave equation in a three-dimensional geometry proceeds as in Paragraph~\ref{III.2}, where the contributions to the charge density from the resonant and non-resonant electrons are calculated separately. 

The expression for the charge density induced by the resonant electrons is a plain generalization of the 1-D result Eq.~(\ref{eq32}), namely, 
\begin{eqnarray}
\nonumber
\rho_w^{res}\approx &&e\iint \left( \int_{v_\phi-\Delta V_R}^{v_\phi-\Delta V_R} \delta \mx.\partial_{\mx} \left[f_H(\mx,v_\phi,\mw,t)+(V_k-v_\phi)\partial_{V_k}f_H\vert_{v_\phi}\right]dV_k\right)d\mw\\
\label{eq67}
&&+ e\iint \left( \int_{v_\phi-\Delta V_R}^{v_\phi+\Delta V_R}\mv. \partial_{\mV}f_H\vert_{v_\phi,\mw}dV_k \right) d\mw,
\end{eqnarray}
where $V_k$ is the projection of $\mV$ along the local direction of the wave vector, and $\mw$ is perpendicular to that direction. The derivation of $\rho_w^{res}$ is detailed in the Appendix~\ref{B}, where it is found that the value of $\Xi^{res}$ is exactly the same as in 1-D, namely,
\begin{equation}
\label{eq68}
\Xi^{res} \approx -i\left[1-\frac{i}{k^2}Ê\frac{\partial k}{\partial x_k}+\frac{i}{k}Ê\frac{\partial E_k}{\partial x_k}\right] \frac{\pi e^2}{\varepsilon_0mk^2}f'_H(\mx,v_\phi,t),
\end{equation}
where, $x_k$ and $E_k$ are, respectively, the local direction and the component of the field amplitude along $\mk$, and where, now, we have denoted, $f'_H \equiv \iint \partial_{V_k}f_H d\mathbf{V}_\bot$. 

As regards the contribution from the non-resonant electrons, it is expressed in terms of $\chi$, which is defined as in 1-D,
\begin{eqnarray}
\nonumber
\chi& = &-\frac{e^2}{\varepsilon_0mk} \int_{-\infty}^{+\infty}Ê\frac{f^{'nres}_H}{kV_k-\omega}dV_k \\
\label{eq69}
& \approx &  -\frac{e^2}{\varepsilon_0mk} P.P.\left( \int \frac{f^{'}_H}{kV_k-\omega}dV_k\right).
\end{eqnarray}
Then, as shown in the Appendix~\ref{A}, the real and imaginary parts of $\Xi^{nres}$ are, respectively,
\begin{eqnarray}
\label{eq70}
\Xi^{nres}_r&=&\chi, \\
\label{eq71}
\Xi^{nres}_i&=&Ê\partial^{2}_{t\omega}\chi-\mathbf{\nabla}.\partial_{\mk}\chi+\frac{\mk.\left[Ê\partial_\omega \chi \partial_t \mE-(\partial_\mk \chi.\mathbf{\nabla}) \mE\right]-\mE.\mathbf{\nabla}\chi-\chi \mathbf{\nabla}.\mE }{\mk.\mE}.
\end{eqnarray}
Then, Eq.~(\ref{II7}) from Gauss law, $1+\Xi_r=0$, yields,
\begin{equation}
\label{eq72}
\chi = -1-\frac{1}{k^2}Ê\frac{\partial k}{\partial x_k}+\frac{1}{k}Ê\frac{\partial E_k}{\partial x_k},
\end{equation}
so that, at first order in the space variations of the fields, $\mathbf{\nabla}\chi \approx 0$ and  $-\chi \mathbf{\nabla}.\mE  \approx \mathbf{\nabla}.\mE$. Moreover, we make use of the same approximation as in 1-D and replace, in the expression for $\Xi_i$, $\mE$ by $\mathbf{E}_p$. Then, Eq.~(\ref{II8}) from Gauss law, $\Xi_i-\mathbf{\nabla}.\mathbf{E}_p/\mathbf{k}.\mathbf{E}_p=\mk.\mathbf{E}_d \cos(\delta \phi)/{\mk.\mathbf{E}_p}$, reads,
\begin{equation}
\label{eq73}
\frac{\partial^{2}_{t\omega}\chi-\mathbf{\nabla}.\partial_{\mk}\chi}{2}+\frac{\mk.\left[Ê\partial_\omega \chi \partial_t \mathbf{E}_p-(\partial_\mk \chi.\mathbf{\nabla}) \mathbf{E}_p\right]}{\mk.\mathbf{E}_p}-\frac{\pi e^2}{\varepsilon_0mk^2}f'_H(\mx,v_\phi,t)=\frac{\mk.\mathbf{E}_d}{\mk.\mathbf{E}_p} \cos(\delta \phi).
\end{equation}
Now, for a plasma wave, up to terms which are of order one in its space variations, $\mathbf{E}_p \approx E_k(\mk/k)$ (see, for example, Ref.~\cite{srs3D}). Hence, up to terms which are of order two in the fields variations, one may use in  Eq.~(\ref{eq73}), $\mathbf{E}_p \approx E_k(\mk/k)\approx E_p(\mk/k)$, which yields,
\begin{equation}
\label{eq74}
\frac{\partial}{\partial t}\left( \frac{\partial \chi}{\partial \omega}E_p^2\right)-\mathbf{\nabla}.\left( \frac{\partial \chi}{\partial \mk}E_p^2\right) +2\nu_L \frac{\partial \chi}{\partial \omega}E_p^2=2E_pE_d\cos(\delta \phi),
\end{equation}
where the Landau damping rate, $\nu_L$, is still given by Eq.~(\ref{eq46}) or (\ref{eq47}). Hence, one recovers the 1-D result that, if the wave is not driven, its action decays at twice the Landau damping rate. 
\section{Conclusion}
\label{V}
Starting from first principles, we described, in this paper, the linear propagation of an electron plasma wave. Our analysis accounted for any type of force field that might make the plasma evolve, in space and time, on hydrodynamical scales. Therefore, our main assumption was to consider the plasma as collisionless. However, we also had to exclude the very special situation when there was, in the plasma, an extremely cold beam whose mean speed was the wave phase velocity. Except for the two former hypotheses, our analysis was totally general, and came to the very simple conclusion that, if the wave was not driven, its action decayed at the Landau damping rate, $\nu_L$. Moreover, we found no correction to the WKB expression for $\nu_L$, at first order in the variations of $\mk$ and $\omega$. Therefore, we showed that the concept of wave action was very robust, valid in the most general situation. Moreover, we provided a solid theoretical basis to the simple use of a WKB formula in order to predict the evolution of the wave action, a result that was not completely obvious, although abundantly used.

When deriving the Landau damping rate, in the envelope equation for the EPW, we showed that the electrons that mainly contributed to it were those whose velocity lied in the range, $[v_\phi-\Delta V,v_\phi+\Delta V]$, where $\Delta V$ was time-dependent. For short times, $\Delta V \sim 1/kt$, until it reached $\max(2\pi/kT_0, 2\pi v_\phi/kL_0)$, $T_0$ and $L_0$ being, respectively, the typical time and length scales of variation of the EPW amplitude. This extended previous results on the resonance width for Landau damping~\cite{drummond, gris} to an EPW amplitude that may vary in space and time, and that does not necessarily decays.

Our analysis of the EPW propagation was made general enough to account for an external drive, which was necessary to be able to address the important issue of stimulated Raman scattering, that motivated this work. Actually, the present paper is the cornerstone of forthcoming articles that discuss how to model SRS by accounting for kinetic effects in an inhomogeneous and non stationary plasma, in the linear and non linear regimes.

As for the linear propagation of an EPW, this is, clearly, one of the oldest issue in plasma physics, if not the oldest one.  As stated in the introduction, it was not obvious at all that, for any kind of plasma, one could simply model it by the blind application of Whitham's theory, completed by the blind use of a WKB formula for the Landau damping rate. This result is now proved.
\begin{acknowledgments}
The author acknowledges I.Y. Dodin for useful discussions.  
\end{acknowledgments} 

\appendix
\section{Derivation of $\bm{\Xi}$ when there is no resonant electron}
\label{A}
\setcounter{equation}{0}
\newcounter{app}
\setcounter{app}{1}

\label{A}
\subsection{One-dimensional geometry}
In order to derive the electron charge density, $\rho_w$, given by Eq.~(\ref{eq12}), we use the results of Section~\ref{III.1}, that $\delta v = \delta v_0e^{i\varphi}+c.c.$ and  $\delta x = \delta x_0e^{i\varphi}+c.c.$, with $\delta v_0$ and $\delta x_0$ respectively given by Eqs.~(\ref{eq25}) and~(\ref{eq26}). Then, from Eq.~(\ref{eq13}) is obtained an expression for $\Xi$, in terms of the electron susceptibility, $\chi$, which, when there is no resonant electron, is
\begin{equation}
\label{A1}
\chi = -\frac{e^2}{\varepsilon_0mk} \int_{-\infty}^{+\infty}Ê\frac{f'_H}{kV_0-\omega}dV_0.
\end{equation}

Now, from the expression Eq.~(\ref{eq26}) of Paragraph~\ref{II.1} for $\delta x$, and by making use of an integration by parts, one finds,
\begin{equation}
\label{A7}
e \int_{-\infty}^{+\infty} \delta x \partial_x f_H(x,V_0,t)dV_0=\frac{e^2E_0e^{i\varphi}}{mk}\int_{-\infty}^{+\infty} \frac{\partial_xf'_H}{kV_0-\omega}dV_0+c.c.
\end{equation}
Using Eq.~(\ref{A7}), together with the expressions Eq.~(\ref{eq25}) for $\delta v$, Eq.~(\ref{eq12}) for $\rho_w$ yields, $\rho_w=\rho_0e^{i\varphi}+c.c$, with
\begin{eqnarray}
\nonumber
\frac{m}{e^2}\rho_0=&& iE_0 \int_{-\infty}^{+\infty} \frac{f'_H}{kV_0-\omega}dV_0+\frac{E_0}{k}\int_{-\infty}^{+\infty} \frac{\partial_xf'_H}{kV_0-\omega}dV_0 \\
\nonumber
&&+ \partial_tE_0\int_{-\infty}^{+\infty}\frac{f'_H}{(kV_0-\omega)^2}dV_0+\partial_xE_0 \int_{-\infty}^{+\infty}\frac{V_0f'_H}{(kV_0-\omega)^2}dV_0\\
\label{A8}
&&+E_0  \int_{-\infty}^{+\infty}f'_H\frac{-\partial_t \omega+kF_H/m+2V_0 \partial_t k/\partial t+V_0^2 \partial^2k/\partial x^2}{(kV_0-\omega)^3} dV_0.
\end{eqnarray}
We now take advantage of the fact that $f_H$ obeys the Vlasov equation~(\ref{eq6}) to find, $(F_H/m)f'_H=-\partial_tf_H-V_0\partial_xf_H$. This yields, by making use of an integration by parts, 
\begin{equation}
\label{A9}
 \int_{-\infty}^{+\infty}\frac{k(F_H/m)f'_H}{(kV_0-\omega)^3} dV_0=-\frac{1}{2} \int_{-\infty}^{+\infty} \frac{\partial_tf'_H+V_0\partial_xf'_H}{(kV_0-\omega)^2}-\frac{1}{2k} \int_{-\infty}^{+\infty}\frac{\partial_xf'_H}{kV_0-\omega}dV_0.
\end{equation}
Therefore, from Eqs.~(\ref{A8}) and (\ref{A9}), one easily finds that $\Xi \equiv i\rho_0/\varepsilon_0kE_0$ is,
\begin{eqnarray}
\nonumber
\Xi=&&\chi+\frac{i}{E_0}\frac{\partial \chi}{\partial \omega}\frac{\partial E_0}{\partial t}-\frac{i}{E_0}\left(\frac{\partial \chi}{\partial k}+\frac{\chi}{k}\right)\frac{\partial E_0}{\partial x}\\
\nonumber
&&+\frac{i}{2}Ê\left[\frac{\partial^2 \chi}{\partial \omega^2}\frac{\partial \omega}{\partial t}+\frac{\partial^2\chi}{\partial k \partial \omega} \frac{\partial k}{\partial t}-\frac{e^2}{\varepsilon_0mk} \int_{-\infty}^{+\infty} \frac{\partial_tf'_H}{(kV_0-\omega)^2}\right] \\
\nonumber
&&-\frac{i}{2}Ê\left[\frac{\partial^2 \chi}{\partial k^2}\frac{\partial k}{\partial x}+\frac{\partial^2\chi}{\partial k \partial \omega}Ê\frac{\partial \omega}{\partial x}+\frac{e^2}{\varepsilon_0mk} \int_{-\infty}^{+\infty} \partial_xf'_H \left\{ \frac{V_0}{(kV_0-\omega)^2}+\frac{1}{k(kV_0-\omega)} \right\}\right]\\
\label{A10}
&&-\frac{i}{k}\left[\frac{\partial \chi}{\partial k}\frac{\partial k}{\partial x}+\frac{\partial \chi}{\partial \omega}\frac{\partial \omega}{\partial x}-\frac{e^2}{\varepsilon_0mk} \int_{-\infty}^{+\infty}\frac{\partial_xf'_H}{kV_0-\omega}\right],
\end{eqnarray}
where we made use of the identity $\partial_tk=-\partial_x\omega$. 

Now, it is clear that the second line of the right-hand side of Eq.~(\ref{A10}) is just $(i/2)\partial_t(\partial \chi/\partial \omega)$, while the third and fourth lines of the right-hand side of Eq.~(\ref{A10}) are, respectively, $-(i/2)\partial_x(\partial \chi/\partial k)$ and $-(i/k)\partial_x \chi$. This straightforwardly yields,
\begin{eqnarray}
\label{A11}
\Xi_r&=&\chi,\\
\label{A12}
\Xi_i&=&\frac{1}{2E_0^2}\left[\frac{\partial}{\partial t}\left( \frac{\partial \chi}{\partial \omega}E_0^2\right)-\frac{\partial}{\partial x}\left( \frac{\partial \chi}{\partial k}E_0^2\right) \right]-\frac{\chi}{kE_0}\frac{\partial E_0}{\partial x}-\frac{1}{k}\frac{\partial \chi}{\partial x},
\end{eqnarray}
where $\Xi_i \equiv \Im(\Xi)$ and $\Xi_r \equiv \Re(\Xi)$.

\subsection{Three-dimensional geometry}
In a three-dimensional geometry, we find it more convenient to make use of an integration by parts in the second term of Eq.~(\ref{II4}) to find
\begin{equation}
\label{A13}
\rho_w^{nres}(\mx,t) = e \iiint\delta \mx.\partial_\mx f_H^{nres}(\mx,\mV,t)d\mV- e\iiint \sum_n\frac{\partial\delta v_n}{ \partial V_{0n}}f_H^{nres}(\mx,\mV,t)d\mV,
\end{equation}
where $\delta v_n$ and $V_{0n}$ are, respectively, the $n^{th}$ component of $\mv$ and $\mV$. Similarly, by making use of an integration by parts, we find,
\begin{equation}
\label{A14}
\chi = -\frac{e^2}{\varepsilon_0m} \iiintÊ\frac{f_H}{(kV_0-\omega)^2}dV_0.
\end{equation}
Recalling that the total electric field is, $\mathbf{E}=\mE e^{i\varphi}+c.c.$, and using the same method as in Paragraph~\ref{III.1}, one easily finds that, to the relevant orders, $\mathbf{\delta x} =\mathbf{\delta x}_0e^{i\varphi}+c.c.$ and $\mathbf{\delta v}=\mathbf{\delta v}_0e^{i\varphi}+c.c.$ with,   
\begin{eqnarray}
\label{A15}
\mathbf{\delta x_0} = &&\frac{e\mE}{m(\mk.\mV-\omega)^2},\\
\nonumber
\frac{m\mv_0}{e}  = && \mE Ê\left\{\frac{i}{\mk.\mV-\omega}-\frac{1}{Ê(\mk.\mV-\omega)^3} \left[\partial_t\omega-\mV.\mathbf{\nabla}(\mk.\mV)-2\mV.\partial_t\mk-\mk.\mathbf{F}_H/m \right]\right\} \\
\label{A16}&&- \frac{\partial_t\mE+(\mV.\mathbf{\nabla})\mE}{(\mk.\mV-\omega)^2}.
\end{eqnarray}
Now, even if the plasma is magnetized, the $n^{th}$ component of $\mathbf{F}_H$ does not depend on the $n^{th}$ component of $\mV$. Therefore,
\begin{eqnarray}
\nonumber
\frac{m}{e}\sum_n \frac{\partial \delta v_{0n}}{\partial V_{0n}} = && \mk.\mE \left\{\frac{-i}{(\mk.\mV-\omega)^2}+\frac{3}{(\mk.\mV-\omega)^4}\left[\partial_t \omega-\mV.\mathbf{\nabla}(\mk.\mV)-2\mV.\partial_t\mk-\mk.\mathbf{F}_H/mÊ\right]Ê\right\}\\
\label{A17}
&&+2 \frac{\mk. \left[\partial_t \mE+(\mV.\mathbf{\nabla}).\mE\right]+\mE.\partial_t\mk+\mE.\mathbf{\nabla}(\mk.\mV)}{(\mk.\mV-\omega)^3}-\frac{\mathbf{\nabla}.\mE}{(\mk.\mV-\omega)^2}.
\end{eqnarray}
By making use of an integration by parts along $V_k$ (the component of $\mV$ along $\mk$), and taking advantage of the fact that $\mk.\mathbf{F}_H$ does not depend on $V_k$, one finds
\begin{equation}
\label{A18}
\iiint \frac{3\mk.\mathbf{F}_H}{m(\mk.\mV-\omega)^4}f_Hd\mV=\iiint  \frac{\mk.\mathbf{F}_H}{m}\frac{\partial_{V_k}f_H}{k(\mk.\mV-\omega)^3}d\mV.
\end{equation}
From the Vlasov equation~(\ref{II1}), 
\begin{equation}
\label{A18b}
 \frac{\mk.\mathbf{F}_H}{m}\partial_{V_k}f_H=-k\left[ \frac{\partial f_H}{\partial t}+(\mV.\mathbf{\nabla})f_H-\frac{\mathbf{F}_\bot}{m}\frac{\partial f_H}{\partial \mathbf{V}_\bot}\right],
\end{equation}
where $\mathbf{F}_\bot$ and $\mathbf{V}_\bot$ are, respectively, the components of $\mathbf{F}_H$ and $\mV$ transverse to $\mk$. Then, since $\mathbf{F}_\bot$ does not depend on $\mathbf{V}_\bot$, from Eqs.~(\ref{A18}) and (\ref{A18b}), 
\begin{equation}
\label{A19}
\iiint \frac{3\mk.\mathbf{F}_H}{m(\mk.\mV-\omega)^4}f_Hd\mV=-\iiint  \frac{\partial_t f_H+(\mV.\mathbf{\nabla})f_H}{(\mk.\mV-\omega)^3}d\mV.
\end{equation}
Moreover, taking advantage of the identity $\partial_t \mk=-\mathbf{\nabla}Ê\omega$, one finds,
\begin{eqnarray}
\nonumber
\frac{e^2\mE}{m}. \iiint \left\{ 2f_H \left[ \frac{\partial_t\mk-\mathbf{\nabla}(\mk.\mV)}{(\mk.\mV-\omega)^3}Ê\right]+\frac{\mathbf{\nabla}f_H}{(\mk.\mV-\omega)^2} \right\}d\mV&=&\mE.\mathbf{\nabla}\frac{e^2}{m}\iiint \frac{f_H}{(\mk.\mV-\omega)^2}d\mV\\
\label{A20}
&=&-\varepsilon_0\mE.\mathbf{\nabla \chi}.
\end{eqnarray}
Therefore, plugging Eqs.~(\ref{A15}) and~(\ref{A16}) into the definition Eq.~(\ref{A13}) for $\rho_w^{nres}$, and using the identities Eqs.~(\ref{A19}) and~(\ref{A20}), one finds that $\rho_w^{nres}$ reads, $\rho_w^{nres}=\rho_0e^{i\varphi}+c.c.$, with
\begin{equation}
\label{A21}
\frac{\rho_0}{\varepsilon_0 }=\frac{\mk.\mE}{2} \left[Ê\partial^{2}_{t\omega}\chi-\mathbf{\nabla}.\partial_{\mk}\chi\right]+\mk.\left[Ê\partial_\omega \chi \partial_t \mE-(\partial_\mk \chi.\mathbf{\nabla}) \mE\right]-\mE.\mathbf{\nabla}\chi-\chi \mathbf{\nabla}.\mE -i\chi \mk.\mE.\\
\end{equation}
Hence, the real and imaginary parts of $\Xi^{nres} \equiv i\rho_0/\varepsilon_0\mk.\mE$ are, respectively,
\begin{eqnarray}
\label{A22}
\Xi^{nres}_r&=&\chi, \\
\Xi^{nres}_i&=&Ê\frac{\partial^{2}_{t\omega}\chi-\mathbf{\nabla}.\partial_{\mk}\chi}{2}+\frac{\mk.\left[Ê\partial_\omega \chi \partial_t \mE-(\partial_\mk \chi.\mathbf{\nabla}) \mE\right]-\mE.\mathbf{\nabla}\chi-\chi \mathbf{\nabla}.\mE }{\mk.\mE}.
\end{eqnarray}

\section{Definition of the resonant and non-resonant electrons}
\label{C}
\setcounter{equation}{0}
\setcounter{app}{3}
In Section~\ref{II}, the non-resonant electrons are defined to be such that $\vert V_0-v_\phi \vert >\Delta V_R$. This definition needs to be global, i.e., for the non-resonant electrons, $V_0 \in (-\infty, V_1) \cup (V_2, +\infty)$, where $V_1$ and $V_2$ are fixed, independent of space and time, and such that, whatever the local value of $v_\phi$, $v_\phi-V_1>\Delta V_R$ and $V_2-v_\phi > \Delta V_R$. Indeed, if we defined the non-resonant electrons such that $V_0 \in (-\infty, v_\phi-\Delta V_R) \cup (v_\phi+\Delta V_R, +\infty)$ then, the boundaries of the integral in Eq.~(\ref{eq38}) for $\chi$ would depend on $v_\phi$. Consequently, the derivatives of $\chi$ with respect of $k$ and $\omega$ would involve the derivatives of these boundaries, thus affecting the results derived in Paragraph~\ref{III.1}. 

Using a global definition for the non-resonant electrons entails that, for each value of $v_\phi$, the velocity $V_0$ for the resonant electrons is such that $V_0 \in [v_\phi-\Delta V_1,v_\phi+\Delta V_2]$, where $\min(\Delta V_1,\Delta V_2)>\Delta V_R$. Then, from Eq.~(\ref{eq32.2}),  the contribution, $\rho_w^{res}$, to the charge density from the resonant electrons writes $\rho_w^{res}=\rho_0^{res}e^{i\varphi}+c.c.$, with
\begin{equation}
\label{C1}
\rho_0^{res}\approx -\frac{e^2}{m}f'_H(x,v_\phi,t) \int_0^t\int_{-\infty}^{+\infty} \tilde{E}_\kappa(t')e^{i\kappa x}\left( \int_{v_\phi-\Delta V_1}^{v_\phi+\Delta V_1}+ \int_{v_\phi+\Delta V_1}^{v_\phi+\Delta V_2}\right) e^{i[(k+\kappa)V_0-\omega](t'-t)}dV_0d\kappa dt'.
\end{equation}
The first term in Eq.~(\ref{C1}), involving the velocity integral from $v_\phi-V_1$ to $v_\phi+V_1$, clearly yields the value for $\rho_0^{res}$ derived in Paragraph~\ref{III.2.2} since $\Delta V_1>\Delta V_R$. As for the second term, involving the velocity integral from $v_\phi+V_1$ to $v_\phi+V_2$, it is proportional to
\begin{eqnarray}
\nonumber
J_2 &\equiv &\int_0^t\int_{-\infty}^{+\infty} \tilde{E}_\kappa(t')e^{i\kappa x}\int_{v_\phi+\Delta V_1}^{v_\phi+\Delta V_2} e^{i[(k+\kappa)V_0-\omega](t'-t)}dV_0d\kappa dt' \\
\label{C2}
&=& e^{ik\Delta V_m t}\int_0^t\int_{-\infty}^{+\infty} e^{-ik\Delta V_m t'}\tilde{E}_\kappa(t')e^{i\kappa x} \frac{2\sin k\Delta V(t'-t)}{t'-t}d\kappa dt',
\end{eqnarray}
with  $\Delta V_m \equiv (\Delta V_1+\Delta V_2)/2$ and $\Delta V \equiv (\Delta V_1-\Delta V_2)/2$. Eq.~(\ref{C2}) shows that $J_2$ oscillates with time at a period close to $T \equiv 2\pi/k\Delta V_m$. Therefore, the value for $\rho_w^{res}$ derived in Paragraph~\ref{III.2.2} may be viewed as the averaged value of $\rho_w^{res}$ over $T$, so that the envelope equation~(\ref{eq45}) is also an averaged equation over $T$. However, since $\Delta V_m > \Delta V_R$, and since $\Delta V_R$ is chosen so that $k\Delta V_R \gg T_0$, where $T_0$ is the typical time of variation of $E_0$, $k$ or $\omega$, an envelope equation averaged over $T=2\pi/k\Delta V_m$ may be considered as the instantaneous one.

Note, moreover, that if $\Delta V \ll \Delta V_m$, $J_2$ is of the order of $E_0 \times \max(\Delta V/\Delta V_m, T_0/k\Delta V_m)$. In this case, the correction to $\rho_w^{res}$ is negligible at any time, and not only on the average. \\

Now, it may happen that the phase velocity is so different in two remote locations that the conditions $\vert V_0-v_\phi \vert>\Delta V_R$ over the whole plasma, and $\Delta V_R \ll v_T$, may not be both fulfilled. In this case, we simply divide the plasma into domains of finite length. These are such that the  conditions $\vert V_0-v_\phi \vert>\Delta V_R$ over the whole domain, and $\Delta V_R \ll v_T$ may be both fulfilled. Then, for each domain, the envelope equation~(\ref{eq45}) is valid, with $\chi$ very close to the value given by Eq.~(\ref{eq39}). Hence, the same envelope equation holds over each domain, so that this equation is valid for the whole plasma. 

\section{Derivation of $\bm{\Xi^{res}}$ for a nonstationary and inhomogeneous plasma}
\label{B}
\setcounter{equation}{0}
\setcounter{app}{3}
\subsection{One-dimensional geometry}
In 1-D, the value of $\Xi^{res}$ is derived from that of the charge density, $\rho_w^{res}$, induced by the resonant electrons, and given by Eq.~(\ref{eq32}) of Paragraph~(\ref{III.2.2}), 
\begin{equation}
\label{B1}
\rho_w^{res}\equiv \rho_1+\rho_2,
\end{equation}
where 
\begin{eqnarray}
\label{B2}
\rho_1 \approx &&e \int_{v_\phi-\Delta V_R}^{v_\phi-\Delta V_R} \delta x \partial_x \left[f_H(x,v_\phi,t)+(V_0-v_\phi)f'_H(x,v_\phi,t)\right]dV_0,\\
\label{B3}
\rho_2 \approx && e\int_{v_\phi-\Delta V_R}^{v_\phi+\Delta V_R}\delta v  f'_H(x,v_\phi,t)dV_0.
\end{eqnarray}
Let us start by estimating the term $\rho_1$. Since we restrict to first order equations in space and time, we only need to express $\delta x$ at zero order in the variations of $E_0$, $k$ and $\omega$. Moreover, the force $F_H$ may be neglected in the electrons dynamics, so that, using the notations $\cE_0(t) = E_0[x(t),t]$, one finds,
\begin{equation}
\label{B4}
\delta x \approx \frac{-e}{m} e^{i(kx-\omega t)} \int_0^t\int_0^{t'} \cE_0(t'')e^{i(kV_0-\omega)(t''-t)}dt''dt'+c.c.
\end{equation}
Therefore, using the fact that $1/k \Delta V_R$ is much less than the typical time of variation of $\cE_0$, one finds that, when $\Delta V_R t \gg 1$,
\begin{eqnarray}
\nonumber
\int_{v_\phi-\Delta V_R}^{v_\phi-\Delta V_R} \delta x dv_0 \approx   &&\frac{-2ie}{m} e^{i(kx-\omega t)} \int_0^t\int_0^{t'} \cE_0(t'') \frac{\sin[k\Delta V_R (t-t'')]}{k (t-t'')} dt'dt''+c.c. \\
\nonumber
\approx && \frac{-i\pi e}{m} e^{i(kx-\omega t)} \int_0^t\int_0^{t'} \cE_0(t'') \delta (t-t'') dt'dt''+c.c. \\
\label{B5}
=&&0.
\end{eqnarray}
Similarly, 
\begin{eqnarray}
\nonumber
\int_{v_\phi-\Delta V_R}^{v_\phi-\Delta V_R} (V_0-v_\phi) \delta x dv_0 \approx   &&\frac{-2ie}{m} e^{i(kx-\omega t)} \int_0^t\int_0^{t'} \frac{\cE_0(t'')}{ik}Ê\partial_{t''} \left( \int_{v_\phi-\Delta V_R}^{v_\phi-\Delta V_R}e^{i(kV_0-\omega)(t''-t)} dV_0 \right) dt'dt'' \\
\nonumber
\approx && \frac{-\pi e}{m} e^{i(kx-\omega t)} \int_0^t\int_0^{t'} \frac{\cE_0(t'')}{k} Ê\partial_{t''}\delta (t-t'') dt'dt''+c.c. \\
\label{B6}
=&&0.
\end{eqnarray}
We, therefore, conclude that, at first order in the fields variations, $\rho_1 = 0$. \\

Let us now estimate $\rho_2$, using the same method as in Subsection~\ref{III.2.2}. Hence, we start by expressing $E_0[x(t),t)]$ as a Fourier integral, $E_0(x,t)=\int_{-\infty}^{+\infty} \tilde{E}_\kappa(t)e^{i\kappa x}d\kappa$. Then, since the total field is, $E(x,t)=E_0e^{i\varphi(x,t)}+c.c$, introducing the notation 
\begin{equation}
\label{B6b}
\alpha_\kappa(x,t)\equiv\kappa x+\varphi(x,t),
\end{equation}
one finds,
\begin{equation}
\label{B7}
E(x,t)=\int_{-\infty}^{+\infty} \tilde{E}_\kappa(t)e^{i \alpha_\kappa(x,t)}d\kappa +c.c.
\end{equation}
Since we restrict here to the linear response with respect to $E$, this lets us express the electron position the following way,
\begin{eqnarray}
\nonumber
x(t') \approx &&x(t)+V_0(t'-t)+\int_{t'}^{t}\int_0^{t''}Ê\frac{F_H(t''')}{m}dt'''dt''\\
\label{B8}
\equiv && x(t)+V_0(t'-t)+\delta x_H,
\end{eqnarray}
where $F_H$ is the force term in the Vlasov equation for $f_H$,~Eq.~(\ref{eq6}), accounting for the fact that the plasma is inhomogeneous and nonstationary. 

Hence, from Eqs.~(\ref{B3}),~(\ref{B7}) and~(\ref{B8}), one finds 
\begin{equation}
\label{B10}
\rho_2 = \frac{-e^2}{m}f'_H(x,v_\phi,t) \int_{-\infty}^{+\infty} \int_0^t \tilde{E}_\kappa(t') I_3(t'-t)dt'd\kappa+c.c.,
\end{equation}
where
\begin{equation}
\label{B11}
I_3(t'-t)=\int_{v_\phi-\Delta V_R}^{v_\phi+\Delta V_R} e^{i \alpha_\kappa[x+V_0(t'-t)+\delta x_H,t']}dV_0.
\end{equation}
The integral $I_3(t'-t)$ is calculated exactly the same way as the integral $I_1(t)$ of Paragraph~\ref{II.1}, which yields,
\begin{eqnarray}
\nonumber
I_3(t'-t)& \approx &\left[ e^{i\alpha_\kappa[x+V_0(t'-t)+\delta x_H,t']} \left\{ \frac{-i}{\partial \alpha_\kappa/\partial V_0}-\frac{\partial ^2 \alpha_\kappa/\partial V_0^2}{(\partial \alpha_\kappa/\partial V_0)^3}\right\} \right]_{v_\phi-\Delta V_R}^{v_\phi+\Delta V_R} \\
\nonumber
&=&\frac{-i}{[k(x,t)+\kappa](t'-t)}\left(\mathcal{I}[x+(v_\phi+\Delta V_R)(t'-t)+\delta x_H,t']e^{i\alpha_\kappa[x+(v_\phi+\Delta V_R)(t'-t)+\delta x_H,t']}\right.\\
\label{B12}
&&\left.-\mathcal{I}[x+(v_\phi-\Delta V_R)(t'-t)+\delta x_H,t']e^{i\alpha_\kappa[(v_\phi-\Delta V_R)(t'-t)+\delta x_H,t']} \right),
\end{eqnarray}
where
\begin{equation}
\label{B13}
\mathcal{I}[\zeta,t'] =\frac{k(x,t)+\kappa}{k(\zeta,t')+\kappa}\left[1-i\frac{\partial k/\partial \zeta}{[k(\zeta,t')+\kappa]^2}\right].
\end{equation}
By hypothesis, $\Delta V_R$ is much smaller than the typical scale of variation in velocity, $v_T$, of $f'_H$. Unless for very peculiar distribution functions, this entails $\Delta V_R \ll v_\phi$. Moreover, we also assume that $\delta x_H \ll v_\phi(t-t')$, which would usually be the case  since $\delta x_H$ is due to the space and time variations of the density, which changes on scales much larger than the fields $E_0$, $k$ and $\omega$. Therefore, since $\mathcal {I}$ is a slowly varying function, one may use the approximation, $\mathcal{I}[x+(v_\phi\pm \Delta V_R)(t'-t)+\delta x_H,t'] \approx \mathcal{I}[x+v_\phi(t'-t),t']$. Moreover, we henceforth also neglect $\delta x_H$ in the phase $\alpha_\kappa$. Since this phase varies quickly, its local value does depend on $\delta x_H$, however, because $(v_\phi \pm \Delta V_R)(t'-t)$ varies much more rapidly with $t'$ than $\delta x_H$, the time integral of $I_3(t'-t)$ should not be much affected by the neglect of $\delta x_H$. Hence, we come to the following approximate value for $I_3$, 
\begin{equation}
\label{B14}
I_3(t'-t) \approx-i \frac{\mathcal{I}[x+v_\phi(t'-t),t']}{[k(x,t)+\kappa](t'-t)} \left\{e^{i\alpha_\kappa[x+(v_\phi+\Delta V_R)(t'-t),t']}Ê-e^{i\alpha_\kappa[(v_\phi-\Delta V_R)(t'-t),t']}\right\}.
\end{equation}
Clearly, when $t'\rightarrow t$, $I_3(t'-t) \rightarrow \Delta V_R$, while, when $\vert t'-t \vert >2\pi/(k+\kappa)\Delta V_R$, $I_3(t'-t)$ oscillates with $t'$, the amplitude of the oscillations decreasing as $1/(k+\kappa)(t'-t)$. Then, since $\tilde{E}_\kappa$ only assumes significant values when $\kappa \ll k$, and since $1/k\Delta V_R$ is much less than the typical time of variations of $\tilde{E}_\kappa$,  $I_3(t'-t)$ may be considered as a delta function in the time integral of Eq.~(\ref{B10}). Hence, just as in Section~\ref{III.2.2}, in this time integral, $\tilde{E}_\kappa(t')$ may be replaced by $\tilde{E}_\kappa(t)$. Then, Eq.~(\ref{B10}) translates into, 
\begin{equation}
\label{B15}
\rho_2 \approx  -\frac{\pi e^2}{mk}f'_H(x,v_\phi,t)e^{i\varphi(x,t)}\int_{-\infty}^{+\infty}\frac{\tilde{E}_\kappa(t)e^{i\kappa x}}{1+\kappa/k}\eta_\kappa(x,t)d\kappa+c.c.,
\end{equation}
where
\begin{equation}
\label{B16}
\eta_\kappa(x,t) = \frac{-i}{\pi} \int_0^t\mathcal{I}[x+v_\phi(t'-t),t'] e^{-i[\kappa x+\varphi(x,t)]}\frac{e^{i\alpha_\kappa[x+(v_\phi+\Delta V_R)(t'-t),t']}Ê-e^{i\alpha_\kappa[(v_\phi-\Delta V_R)(t'-t),t']}}{(t'-t)}dt'.
\end{equation}
Since $\kappa \ll k$, we may now approximate Eq.~(\ref{B15}) by its expansion about $\kappa=0$, i.e., make use of the approximations,  $1/(1+\kappa/k)\approx1-\kappa/k$ and $\eta_\kappa\approx \eta+\kappa \partial_\kappa \eta_\kappa \vert_{\kappa=0}$, where we have denoted $\eta \equiv \eta_{\kappa=0}$. This yields,
\begin{equation}
\label{B17}
\rho_2 \approx  -\eta \frac{\pi e^2}{mk}f'_H(x,v_\phi,t)e^{i\varphi(x,t)}\left[E_0(x,t)+\frac{i}{k}Ê\frac{\partial E_0}{\partial x}Ê\right]+c.c.,
\end{equation}
where the term $\partial_x E_0\times \partial_\kappa \eta_\kappa\vert_{\kappa=0} $ has been neglected because it is of order two in the fields variations. As for $\eta \equiv \eta_{\kappa=0}$, using Eqs.~(\ref{B6b}),~(\ref{B13}) and~(\ref{B16}), it reads,
\begin{equation}
\label{B18}
\eta \equiv \eta_1+\eta_2,
\end{equation}
with
\begin{eqnarray}
\label{B19}
\eta_1 =&& \frac{-ik(x,t)}{\pi} \int_0^t e^{-i\varphi(x,t)} \frac{e^{i\varphi[x+(v_\phi+\Delta V_R)(t'-t),t']}-e^{i\varphi[x+(v_\phi-\Delta V_R)(t'-t),t']}}{(t'-t)k[x+v_\phi(t'-t),t']}dt',\\
\label{B20}
\eta_2 = && \frac{-k(x,t)}{\pi} \int_0^t e^{-i\varphi(x,t)} \frac{\partial k}{\partial x}\frac{e^{i\varphi[x+(v_\phi+\Delta V_R)(t'-t),t']}-e^{i\varphi[x+(v_\phi-\Delta V_R)(t'-t),t']}}{(t'-t)k^3[x+v_\phi(t'-t),t']}dt'.
\end{eqnarray}
Clearly, at first order in the variations of $k$ and $\omega$, 
\begin{equation}
\label{B21}
\eta_2 \approx -\frac{i}{k^2}Ê\frac{\partial k}{\partial x}.
\end{equation}
In order to estimate $\eta_1$, we introduce,
\begin{eqnarray}
\nonumber
\delta \varphi_{\pm}(x,t')=&&\varphi[x+(v_\phi \pm \Delta V_R)(t'-t),t']-\varphi(x,t)-k[(v_\phi\pm \Delta V_R)(t'-t)]-\omega(t'-t) \\
\label{B22}
= && \varphi[x+(v_\phi \pm \Delta V_R)(t'-t),t']-\varphi(x,t)\mp k\Delta V_R(t'-t).
\end{eqnarray}
Then, for $\tau$ small enough,
\begin{equation}
\label{B23}
\delta \varphi_{\pm}(t'+\tau) \approx \delta \varphi_\pm(t')+\frac{(v_\phi \pm \Delta V_R)^2\tau^2}{2}Ê\frac{\partial k}{\partial x}-(v_\phi \pm \Delta V_R)\tau^2Ê\frac{\partial \omega}{\partial x}-\frac{\tau^2}{2}Ê\frac{\partial \omega}{\partial t}.
\end{equation}
Using the hypothesis $v_\phi \gg \Delta V_R$, one finds that, when $\tau \sim 1/k\Delta V_R$, 
\begin{eqnarray}
\nonumber
\frac{(v_\phi \pm \Delta V_R)^2\tau^2}{2}Ê\frac{\partial k}{\partial x} &\sim & \frac{v_\phi}{k^2\Delta V_R^2}\frac{\partial k}{\partial x}\\
&=& \frac{\omega}{k \Delta V_R}\frac{1}{k^2}Ê\frac{\partial k}{\partial x}.
\end{eqnarray}
Now, $\Delta V_R$ is chosen so that $\omega/\Delta V_R$ be much larger that the typical length of variation, $L_0$, of $k$. Hence, 
\begin{equation}
\label{B24}
(v_\phi \pm \Delta V_R)^2\tau^2Ê\frac{\partial k}{\partial x} \ll \frac{1}{kL_0}\frac{1}{k^2}\frac{\partial k}{\partial x}.
\end{equation}
Similarly, 
\begin{eqnarray}
\nonumber
(v_\phi \pm \Delta V_R)\tau^2Ê\frac{\partial \omega}{\partial x} &\sim&  \frac{\omega^2}{k^2\Delta V_R^2}\frac{1}{\omega k}\frac{\partial \omega}{\partial x} \\
\label{B25}
& \ll& \frac{1}{(kL_0)^2}\frac{1}{\omega k}\frac{\partial \omega}{\partial x},
\end{eqnarray}
and
\begin{eqnarray}
\nonumber
\tau^2\frac{\partial \omega}{\partial t} &\sim & \frac{\omega^2}{k^2\Delta V_R^2}\frac{1}{\omega^2}\frac{\partial \omega}{\partial t}Ê\\
\label{B26}
& \ll &\frac{1}{(kL_0)^2}\frac{1}{\omega^2}\frac{\partial \omega}{\partial t}.
\end{eqnarray}
Therefore, over a time of the order of $1/k\Delta V_R$, $\delta \varphi_{\pm}$ varies in a negligible fashion. This shows that $e^{i\varphi[x+(\varphi\pm \Delta V_R)(t'-t),t']}$ oscillates at a frequency very close to $k\Delta V_R$, so that the integral in the right-hand side of Eq.~(\ref{B19}) does not vary much after $t\agt 2\pi/k\Delta V_R$. In this integral, it is therefore licit to make use of the approximation $\delta \varphi_+ \approx \delta \varphi_- \approx \delta \varphi \equiv \varphi[x+v_\phi(t'-t),t']-\varphi(x,t)$. Actually, from Eqs.~(\ref{B24})-(\ref{B26}), it is clear that the Taylor expansion Eq.~(\ref{B23}) is valid up to a time $\tau$ much larger than $kL_0/k\Delta V_R=L_0/\Delta V_R$. Since $v_\phi \gg \Delta V_R$, over such a large time, it is likely that resonant electrons have travelled over the whole plasma length. Hence, the same Taylor expansion may be used for $\delta \varphi_+$ and $\delta \varphi_-$ so that, indeed, $\delta \varphi_+ \approx \delta \varphi_- \approx \delta \varphi$. We, therefore, obtain the following approximate value for $\eta_1$, 
\begin{equation}
\label{B28}
\eta_1 \approx \frac{k(x,t)}{\pi} \int_0^t e^{-i\delta \varphi(x,t')} \frac{2\sin[k\Delta V_R(t'-t)]}{(t'-t)k[x+v_\phi(t'-t),t']}dt'.
\end{equation}
Let us now introduce $S(t')\equiv (2/\pi)\int_{t'}^t 2\sin[k\Delta V_R(t''-t)]/(t''-t)dt''$. Then, by making use of an integration by parts, one finds,
\begin{equation}
\label{B29}
\eta_1Ê\approx 1-k(x,t)\int_0^t \frac{e^{i\delta \varphi}}{k[x+v_\phi(t'-t),t']}\left[i\frac{d\delta \varphi}{dt'}-\frac{v_\phi \partial_xk+\partial_tk}{k[x+v_\phi(t'-t),t']}Ê\right]ÊS(t')dt'. 
\end{equation}
Now, it is clear that $S(t')$ is a Heaviside-like function that varies from 0 to 1 over a time interval of the order of $1/k\Delta V_R$. Therefore, by making use of a Taylor expansion  for $\delta \varphi$, one finds,
\begin{equation}
\label{B30}
\Im(\eta_1-1) \sim -\tau^2 \left[v_\phi^2 \partial_xk-v_\phi \partial_x \omega-\partial_t \omega Ê\right],
\end{equation}
where $\tau$ is the order of $1/k\Delta V_R$. Since, in this paper, we only account for terms which are, at least, of the order of $k^{-2}Ê\partial_xk$,  $(\omega k)^{-1} \partial_x \omega$, or $\omega^{-2} \partial_t \omega$, from Eqs.~(\ref{B24})-(\ref{B26}), we find $\Im(\eta_1-1)\approx 0$.

Similarly,
\begin{equation}
\label{B31}
\Re(\eta_1-1) \sim \frac{v_\phi \partial_xk+\partial_tk}{k^2\Delta V_R} \approx 0.
\end{equation}

Therefore, from Eqs.~(\ref{B17}), (\ref{B18}) and (\ref{B21}) one finds that, at first order in the fields variations,
\begin{equation}
\label{B32}
\rho_2 \approx - \frac{\pi e^2}{mk}f'_H(x,v_\phi,t)\left[E_0-\frac{iE_0}{k^2}Ê\frac{\partial k}{\partial x}+\frac{i}{k}Ê\frac{\partial E_0}{\partial x}\right]e^{i\varphi(x,t)}+c.c.
\end{equation}
Using Eq.~(\ref{B1}) for $\rho_w^{res}$ and the fact that $\rho_1 \approx 0$, one finds that $\rho_w^{res} \approx \rho_2$ is given by Eq.~(\ref{B32}). Then, from the very definition Eq.~(\ref{eq13}) of $\Xi$,
\begin{equation}
\label{B33}
\Xi^{res} \approx -i\left[1-\frac{i}{k^2}Ê\frac{\partial k}{\partial x}+\frac{i}{k}Ê\frac{\partial E_0}{\partial x}\right] \frac{\pi e^2}{\varepsilon_0mk^2}f'_H(x,v_\phi,t).
\end{equation}

\subsection{Three-dimensional geometry}
The charge density induced by the resonant electrons in a three-dimensional geometry is given by Eq.~(\ref{eq67}) of Section~\ref{IV} and writes
\begin{equation}
\label{B34}
\rho_w^{res} \equiv \rho_1+\rho_2,
\end{equation}
with 
\begin{eqnarray}
\label{B35}
\rho_1\approx &&e\iint \left( \int_{v_\phi-\Delta V_R}^{v_\phi-\Delta V_R} \delta \mx.\partial_{\mx} \left[f_H(\mx,v_\phi,\mw,t)+(V_k-v_\phi)\partial_{V_k}f_H\vert_{v_\phi}\right]dV_k\right)d\mw\\
\label{B36}
\rho_2 \approx&& e\iint \left( \int_{v_\phi-\Delta V_R}^{v_\phi+\Delta V_R}\mv. \partial_{\mV}f_H\vert_{v_\phi,\mw}dV_k \right) d\mw.
\end{eqnarray}
Just like in 1-D, $\rho_1$ is calculated at zero order in the variations of $k$ and $\omega$. Then, the component $\delta x_n$ of $\delta \mx$ is given by Eq.~(\ref{B4}) with $\cE_0$ replaced by $\cE_n$, the component of the field amplitude along the $n^{th}$ direction, and $V_0$ replaced by $V_k$. Then, the integration over $V_k$ straightforwardly yields $\rho_1 \approx 0$. 

 As regards $\rho_2$, in 1-D we found that, except for the term $(1/k^2)(\partial k/\partial x)$ in $\mathcal{I}$, the variations of $k$ and $\omega$ are negligible for the following reasons. i) The velocity integral yields a delta-like function whose temporal width is of the order of $1/k\Delta V_R$. ii) Over this time interval, the resonant electrons move by about $v_\phi/k\Delta V_R$, which is much less than the typical length of variation, $L_0$, of $k$ and $\omega$. Similarly, in 3-D, the integration over $V_k$ will yield a term proportional to $(1/k^2)(\partial k/\partial x_k)$, where $x_k$ is along the local direction of $\mk$. Except for this term, the longitudinal variations of $k$ and $\omega$ are negligible for the same reasons as in 1-D. As for their transverse variations, they are also negligible provided that $V_\bot \alt (kL_0)v_\phi$. If we assume that, for such large values of $V_\bot$ (usually larger than the speed of light), $f_H(\mx,v_\phi,\mw)\approx 0$, then $\mv$ may be considered as a function of $\mk.\mV-\omega$. Consequently, by integration over $\mw$, the contribution to $\rho_2$ from the components of $\mv$ perpendicular to $\mathbf{k}$ yield a negligible contribution. As for the component of $\mv$ along $\mathbf{k}$, its contribution to $\rho_2$ is exactly the same as in 1-D.
 
Therefore, we conclude that the 3-D expression of $\Xi^{res}$ is a mere generalization of Eq.~(\ref{B33}) obtained in 1-D, namely,
\begin{equation}
\label{B37}
\Xi^{res} \approx -i\left[1-\frac{i}{k^2}Ê\frac{\partial k}{\partial x_k}+\frac{i}{k}Ê\frac{\partial E_k}{\partial x_k}\right] \frac{\pi e^2}{\varepsilon_0mk^2}f'_H(\mx,v_\phi,t),
\end{equation}
where, $E_k$ is the projection of the field amplitude along $\mk$, and where, now, we have denoted,
\begin{equation}
\label{B38}
f'_H(\mx,v_\phi,t) \equiv \iint \left. \frac{\partial f_H(\mx,V_k,\mw,t)}{\partial V_k} \right \vert_{V_k=v_\phi} d\mw.
\end{equation}

\end{document}